\documentclass[namedreferences]{solarphysics}

\usepackage[hyperref,optionalrh]{spr-sola-addons} 
\usepackage[optionalrh]{spr-sola-addons} 
\usepackage{graphicx}        
\usepackage{color}           
\usepackage{breakurl}        

\chardef\us=`\_

\begin{document}
	
	\begin{article}
		\begin{opening}
			
			\title{Possible Signature of Sausage Waves in Photospheric Bright Points}
			
			\author[addressref=aff1]{\inits{Y.}\fnm{Yuhang}~\lnm{Gao}\orcid{0000-0002-6641-8034}}
			\author[addressref=aff2,corref,email={lifuyu@ioe.ac.cn}]{\inits{F.}\fnm{Fuyu}~\lnm{Li}\orcid{0000-0002-2569-1632}}
			\author[addressref=aff3]{\inits{B.}\fnm{Bo}~\lnm{Li}\orcid{0000-0003-4790-6718}}
			\author[addressref=aff5]{\inits{W.}\fnm{Wenda}~\lnm{Cao}\orcid{0000-0003-2427-6047}}
			\author[addressref=aff4]{\inits{Y.}\fnm{Yongliang}~\lnm{Song}\orcid{0000-0002-9961-4357}}
			\author[addressref={aff1,aff4}]{\inits{H.}\fnm{Hui}~\lnm{Tian}\orcid{0000-0002-1369-1758}}
			\author[addressref={aff3}]{\inits{M.}\fnm{Mingzhe}~\lnm{Guo}\orcid{0000-0003-4956-6040}}
			\address[id=aff1]{School of Earth and Space Sciences, Peking 
				University, Beijing 100871, People's Republic of China}
			\address[id=aff2]{Institute of Optics and Electronics, Chinese Academy of Sciences, Chengdu, Sichuan, 610209, People's Republic of China}
			\address[id=aff3]{Shandong Provincial Key Laboratory of Optical 
				Astronomy and Solar-Terrestrial Environment, Institute of 
				Space Sciences, Shandong University, Weihai, Shandong 264209, People's 
				Republic of China}
			\address[id=aff5]{Big Bear Solar Observatory, New Jersey Institute of Technology, Big Bear City, CA 92314, USA}
			\address[id=aff4]{Key Laboratory of Solar Activity, National 
				Astronomical Observatories, Chinese Academy of
				Sciences, Beijing 100012, People's Republic of China}

			\runningauthor{Y. Gao et al.}
			\runningtitle{Possible Evidence of Sausage Waves in Photospheric Bright Points}
			
			\begin{abstract}
				Sausage waves have been frequently reported in solar magnetic structures such as sunspots, pores, and coronal loops. However, they have not been unambiguously identified in photospheric bright points (BPs). Using high-resolution TiO image sequences obtained with the \textit{Goode Solar Telescope} at the Big Bear Solar Observatory, we analyzed  four isolated BPs. It was found that their area and average intensity oscillate for several cycles in an in-phase fashion. The oscillation periods range from 100 to 200 \textbf{seconds}. We interpreted the phase relation as a signature of sausage waves, \textbf{particularly} slow waves, after discussing the sausage wave theory and the opacity effect.
			\end{abstract}
		\end{opening}
		
		\section{Introduction}
		\label{intro} 
		
		Photospheric bright points (BPs) are dynamic, small-scale, bright structures in the solar photosphere, usually appearing in dark intergranular lanes.
		They are associated with magnetic elements in magnetograms and have a \textbf{magnetic-field} strength of about \textbf{1\,kG}. 
		BPs are typically circular \citep{berger1995,bovelet2003}, 
		while some of them have elongated shapes \citep{kuckein2019}. 
		Their equivalent diameter is about \textbf{100\,--\,300\,km}, and their lifetime is about \textbf{90\,seconds\,--\,10\,minutes} \citep{berger1996,utz2010,keys2011,keys2014}. 
		In addition, BPs usually have a brightness contrast of \textbf{0.8\,--\,1.8} relative to the mean intensity of the photosphere \citep{almeida2004}.
		
		BPs were first discovered in G-band observations of the photosphere in the 1970s  \citep{Dunn1973}, and later \citeauthor{mehltretter1974observations}  (\citeyear{mehltretter1974observations}) found that they represent \textbf{magnetic-flux} concentrations by comparing Ca \textbf{{\sc ii}} K images with magnetograms. 
		BPs are therefore also called magnetic bright points, and they are considered as the photospheric footpoints of magnetic flux tubes (e.g. \citealp{berger1998,depontieu}). A process named convective collapse is believed to be their formation mechanism \citep{spruit1979}, which has received some observational support
		(e.g. \citealp{solanki1996,rubio2001,nagata2008}).
		
		Since BPs can be regarded as tracers of photospheric magnetic flux tubes, many studies have suggested that BPs may host a variety of MHD waves \citep{gonzalez2011,jess2012,mumford2015,jafarzadeh2017}.
		Given that the dissipation of MHD waves is among the most promising mechanisms for heating the solar corona, it is of great significance to 
		study MHD waves associated with BPs or photospheric magnetic elements.
		So far, possible observational manifestations of kink waves
		\citep{stangalini2013,stangalini2015}
		and Alfv\'{e}n waves \citep{jess2009} have already been found in BPs (or magnetic elements). However, sausage waves have not been unambiguously identified in them. 
		
		Sausage waves (or sausage modes) in a flux tube are characterised by a sausage-like boundary of the tube, first defined by \citeauthor{defouw1976} (\citeyear{defouw1976}).
		A detailed discussion of this wave mode is given in \textbf{Section} \ref{sausage}.
		There are many studies focusing on theoretical understanding
		and seismological applications of sausage modes (e.g. \citealp{chen2015,guo2016}; see also the review by \citealp{li2020}).
		In observations, sausage modes in \textbf{the }photosphere have been widely studied in magnetic pores and sunspots
		(e.g. \citealp{dorotovivc2007,morton2011,grant2015,freij2016}). 
		Also, \citeauthor{morton2012} (\citeyear{morton2012}) found evidence for the simultaneous existence of sausage 
		and kink modes in fine magnetic structures in the chromosphere.
		In the corona, short-period, quasi-periodic pulsations (QPPs) in flares are also generally interpreted as sausage waves (e.g. \citealp{nakariakov2003,van2011,tian2016}; see also the review by \citealp{zimovets2021}).

		\begin{figure}    
			\centerline{\includegraphics[width=\textwidth,clip=]{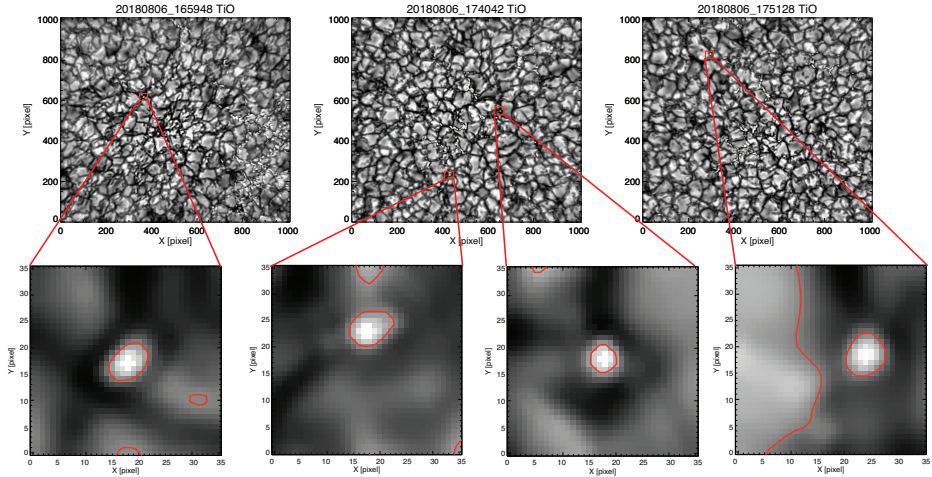}}
			\caption{TiO images with BPs marked by boxes. 
				The images below show \textbf{an enlarged} view of the red box regions, where 
				BPs are identified and tracked with red contours.
			}
			\label{tio}
		\end{figure}
	
		In this \textbf{article}, we report the existence of in-phase oscillations of area and average intensity in BPs using high-resolution TiO images for the first time. Our findings can be regarded as possible evidence of sausage modes in \textbf{photospheric} flux tubes. 
		In fact, magnetoacoustic  waves characterised by intensity or line-of-sight (LOS) velocity oscillations have been found in BPs before \citep{jess2012,jess2012b,stangalini2013b,jafarzadeh2017}.
		But these preivious studies could not tell whether these waves \textbf{were} sausage waves, because either they did not consider BP areas or the area variations were too weak.
		\citeauthor{gonzalez2011} (\citeyear{gonzalez2011}) studied four magnetic patches in circular polarization images obtained by \textbf{the Imaging Magnetograph eXperiment (IMaX) on board \textit{Sunrise}}, and \textbf{they} found that the iso-flux area oscillates with time. 
		However, the time-varying periods suggested that the oscillations may be caused by granular motions rather than a chacteristic oscillation mode. 
		\citeauthor{fujimura2009} (\citeyear{fujimura2009}) also focused on oscillations of intergranular magnetic structures (IMSs) in a magnetogram and interpreted these oscillations as sausage and/or kink waves. But the lifetimes of these IMSs are all \textbf{greater than} one hour, \textbf{which is} much longer than the typical lifetime of BPs. 
		\citeauthor{kolotkov2017} (\citeyear{kolotkov2017}) also detected  long-period (\textbf{80\,--\,230 minutes}) quasi-periodic oscillations of the average magnetic field and total area of a small magnetic \textbf{structure}.
		More recently, \citeauthor{keys2020} (\citeyear{keys2020}) studied the rapid variations of 
		magnetic fields in BPs and believed that such changes are consistent with the behavior of MHD waves, but they did not discuss in detail the changes of intensity.
		
		This \textbf{article} is organised as follows. In Section \ref{obs}, we describe our observation and data analysis methods, 
		and the key results are presented in Section \ref{result}.
		We discuss the \textbf{wave-mode} identification in Section \ref{dis} before giving
		the final conclusion and some remarks in Section \ref{con}.

		\section{Observations and Data Analysis}
		\label{obs}

		The data used in this study were acquired with the 1.6-meter \textit{Goode Solar Telescope} (GST; \citealp{cao2010}) at the Big Bear Solar Observatory (BBSO). We used the Broadband Filter Imager (BFI) installed on \textbf{the} GST to observe the TiO band (7057\,$\mathrm{\mathring{A}}$) from 16:28 to 17:59 UT on 6 \textbf{August} 2018, with a cadence of 15 \textbf{seconds}. The observed regions include the active region
		NOAA 12717 at $ (501^{\prime\prime}, -222^{\prime\prime}) $ from 16:28 to 17:33 and the western lobe of a coronal hole at $(-260^{\prime\prime}, 2^{\prime\prime})$ from 17:33 to 17:59.
		
		The bandwidth of the filter is 10\,$\mathrm{\mathring{A}}$.
		The \textbf{field of view }(FOV) of the TiO data was $ 77^{\prime\prime}\times 77^{\prime\prime} $ and the pixel size was $ 0.034^{\prime\prime} $. 
		During the observation, the seeing condition was better than $ 3^{\prime\prime}$, which is good for the identification of BPs.
		Speckle reconstruction and a high-order adaptive optics (AO) system were used to reach the \textbf{spatial-resolution} limit.
		The data were also destretched to remove the residual \textbf{atmospheric-seeing} effects. 
		
		We first selected isolated BPs in the FOV by \textbf{eye}.
		\citeauthor{liu2018} (\citeyear{liu2018}) and \citeauthor{kuckein2019} (\citeyear{kuckein2019}) divided BPs into isolated ones, which are individual, and non-isolated ones, which display a splitting or merging behavior in observations.
		Obviously, it is more suitable to use isolated BPs to search for MHD waves. \textbf{Because for non-isolated BPs, their splitting and merging behaviors may affect our identification of waves.}
		For the selected BPs, we used contours with a fixed intensity threshold level to identify and track them.
		Then we calculated the area and average intensity of these BPs at each time and examined if there is any oscillation signal.
		Finally, four of the BPs are found to be well isolated and 
		there is also possible evidence of sausage waves inside. 
		Their positions and shapes are shown in Figure \ref{tio}.
		We named them BP1 to BP4. 
		When calculating their average intensity and area, we carefully removed the 
		unrelated contours nearby (we can see them around BP1, BP2, and BP4 in Figure \ref{tio})
		by manually zooming \textbf{on} the region of interest for every frame. We summarize the properties of the four BPs in Table \ref{table1}.
		
		\begin{table}
			\caption{Observational details of the BPs. The area is averaged over the time, and C.C. represents the correlation coeffecient of average intensity and area.}
			\label{table1}
			\begin{tabular}{l c c c c c}     
				\hline                   
				& Date  & Time \textbf{[UT]} & Location &Area \textbf{[Mm$ ^2 $]} & C.C. \\
				\hline
				BP1 & \textbf{06 Aug. 2018} & 16:54:18\,--\,17:03:18 & $(501^{\prime\prime}, -222^{\prime\prime})$ & 0.015  &0.71\\
				BP2 & \textbf{06 Aug. 2018 }& 17:38:12\,--\,17:46:28 &$(-260^{\prime\prime}, 2^{\prime\prime})$ & 0.017 &0.92\\
				BP3 & \textbf{06 Aug. 2018} & 17:33:12\,--\,17:41:27 &$(-260^{\prime\prime}, 2^{\prime\prime})$ & 0.013 &0.69\\
				BP4 & \textbf{06 Aug. 2018} & 17:47:43\,--\,17:53:43 &$(-260^{\prime\prime}, 2^{\prime\prime})$ &0.015 
				&0.89\\
				\hline
			\end{tabular}
		
		\end{table}
		
		\section{Result} 
		\label{result}    
		  
		  \begin{figure}
		  	\centerline{\includegraphics[width=\textwidth,clip=]{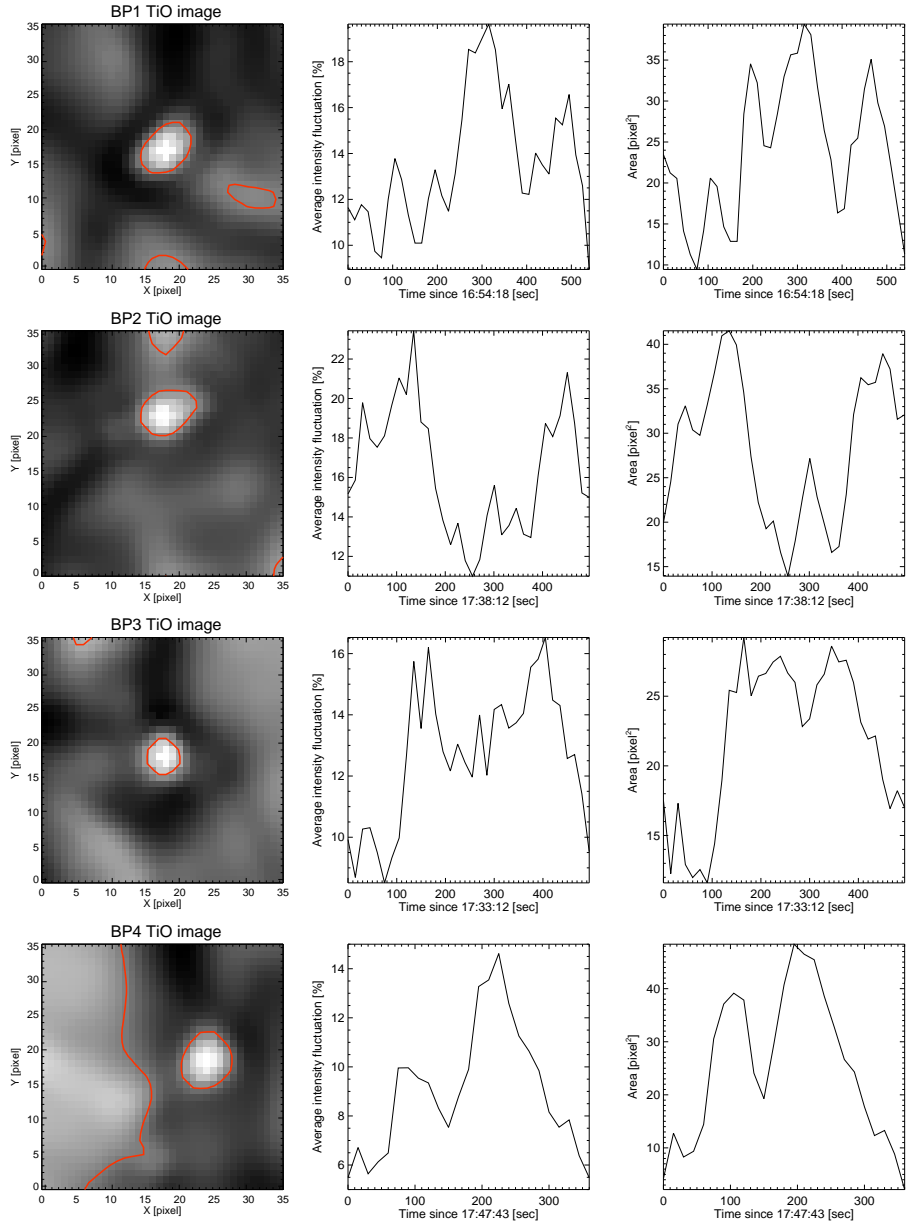}}
		  	\caption{Time series of average intensity and area of the 
		  		four BPs. Each row corresponds to one BP, and the right two columns show the 
		  		temporal variations of the average intensity and area, respectively. Here we use the intensity fluctuation percentage with respect to the mean value of the backgroud.
		  		(\textbf{Animation} of this figure is available \textbf{in the Supplementary Materials.})}     
		  	\label{time}
		  \end{figure}
		
\begin{figure}
			\centerline{\includegraphics[width=\textwidth,clip=]{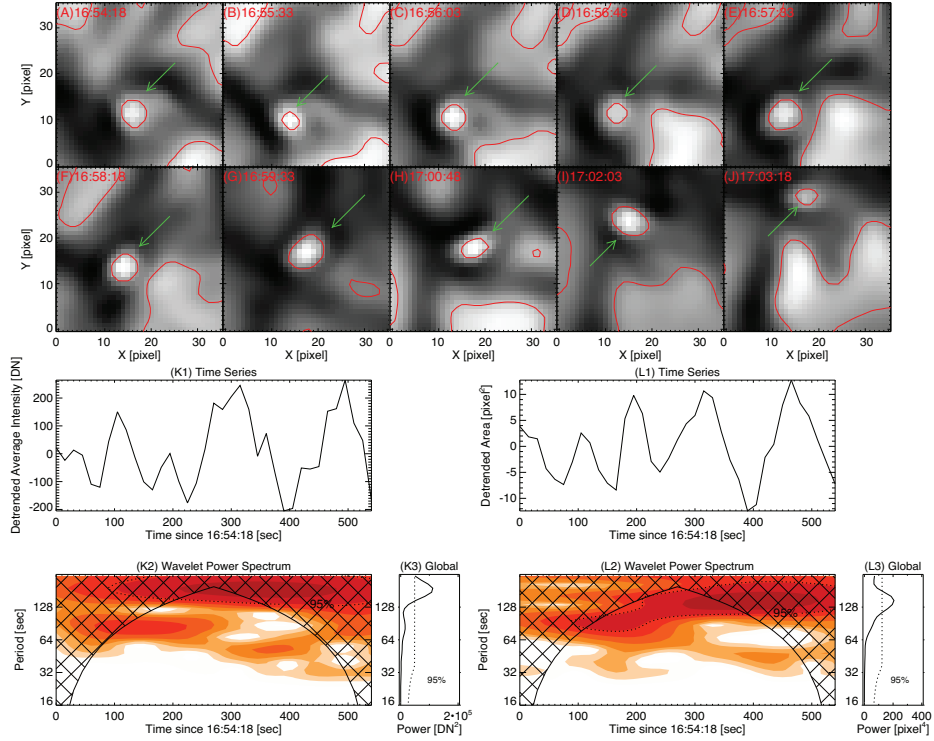}}
			\caption{\textbf{(A\,--\,J)}: snapshots of BP1 marked by green arrows at \textbf{ten} different time. Approximately, every three adjacent panels correspond to an oscillation cycle. (K): wavelet analysis results of detrended time series (shown in Panel K1) of average intensity. The results include the wavelet power spetrum (Panel K2) and the global power spectrum (Panel K3). Darker colors represent stronger powers. The dashed lines indicate a significance level of 95\,\%. (L): same as Panel K but for the detrended time series of area.}     
			\label{case}
		\end{figure}

		Figure \ref{time} presents the time series of the area and average intensity of these four BPs, and we can see obvious oscillations for \textbf{two to four} cycles. 
		It can also be noticed that for each BP, the crests and troughs for these two \textbf{features} appear at nearly the same time, which suggests that the oscillations of area and average intensity are in phase.
		Among these BPs, BP1 has the most oscillation cycles, with four clear crests in the time series of area, respectively at \textbf{105, 195, 315, and 465 seconds}. 
		These instants correspond well to crests in the curve of average intensity.
		\textbf{In addition}, there is also a good correspondence between moments of each trough. 
		Figure \ref{case} gives ten snapshots of BP1 in \textbf{Panels A\,--\,J}, each corresponding to a moment of wave crest or trough. We can clearly see that the BP's area oscillates with time.
		
		Then we detrended the time series of average intensity and area by subtracting a \textbf{three-}minute running average before performing wavelet analysis. The detrended series and wavelet analysis results are shown in Panels K and L of Figure \ref{case}. The peak global wavelet powers for the average intensity and area are at around 150 \textbf{seconds}, which could be seen as the oscillation period. \textbf{For} the other BPs, the periods can be estimated to be \textbf{100\,--\,200 seconds}. The period range we found here are comparable to the results in some previous studies, which interpreted the oscillations they found in BPs as magnetoacoustic waves \citep{jess2012,jess2012b}. We note that oscillations with
		longer periods above \textbf{four minutes} were detected in small magnetic structures before \citep{fujimura2009,gonzalez2011,kolotkov2017}.
		
\begin{figure}
			\centerline{\includegraphics[width=\textwidth]{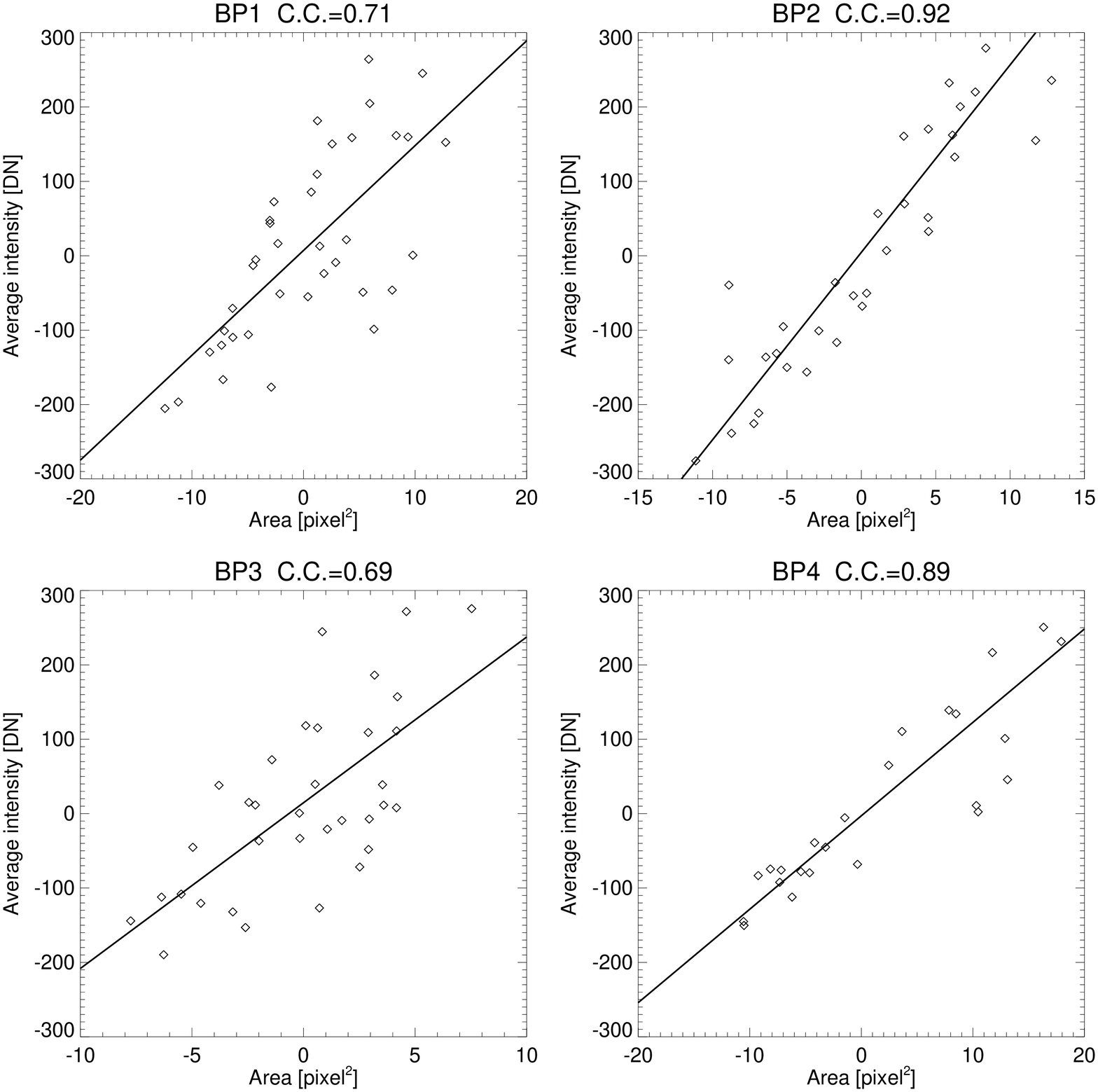}}
			\caption{Scatter plots of the relationship between
				average intensity and area for the four BPs, 
				with solid lines representing linear fits. Note that the average intensity and area are both detrended.}    
			\label{correlation}
		\end{figure}

\begin{figure}
			\centerline{\includegraphics[width=\textwidth]{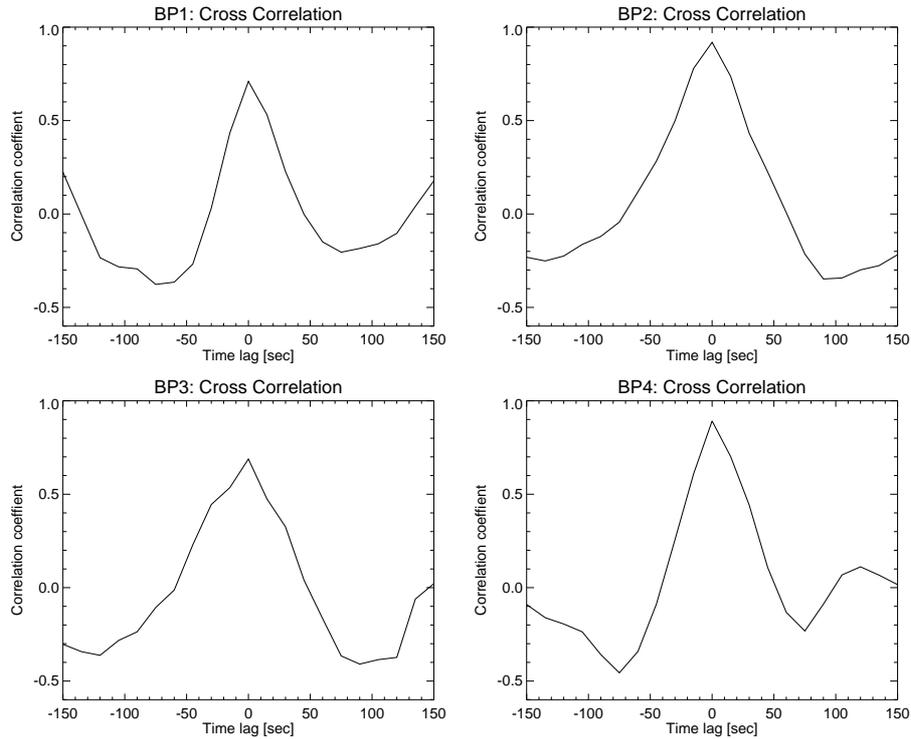}}
			\caption{Lagged cross-correlations between average intensity and area. The
time lag is between the detrended time series of area and 
			  average intensity.}     
			\label{lcc}
		\end{figure}

		In order to further confirm the in-phase relation between oscillations of average intensity and area, we drew scatter plots of relationship between them, and calculated the correlation coefficients, as shown in Figure \ref{correlation}.
		It can be seen that the correlation coefficients are all 
		above 0.68. For BP2 and BP4, the coefficients are around 0.9, showing a good \textbf{positive} correlation.
		Meanwhile, we also performed a lagged cross-correlation (LCC) analysis. From Figure \ref{lcc}, we can see that the coeffients for all BPs reach the peak when time lag is \textbf{zero}. All the results confirm the in-phase relation between average intensity and area.

		A possible question is whether the correlation or the in-phase relation is caused by our analysis method, i.e. using
		contours with a fixed intensity threshold to identify BPs and calculate their parameters. 
		If we consider the total intensity rather than the average intensity in the contours, we might tend to obtain an in-phase relation, since it seems that the total intensity will always increase (or decrease) along with  the increase (or decrease) of the area.
		So here we chose to use time series of the average intensity.
		\textbf{Additionally}, we studied for more than \textbf{four} isolated BPs with the same method, and \textbf{we} found that there are cases without such an oscillation pattern. 
		In conclusion, the in-phase relation is not a result of our \textbf{data-processing} method, but more likely associated with some characteristic oscillation modes.
		
		We will show in the next section that the observed in-phase oscillations
		might be explained as sausage waves in BPs.
	
		\section{Discussion} 
		\label{dis}      
			\subsection{Sausage Mode Waves in a Flux Tube}
		 	\label{sausage}
		 	In the solar atmosphere, magnetic flux tubes are formed with magnetic field lines clumping into tight bundles.
		 	They can be regarded as channels guiding MHD waves. 
		 	A flux tube is usually assumed as a straight cylinder described by cylindrical coordinates $(r, \phi, z)$.
		 	The tube boundary can oscillate in different geometrical patterns, forming various wave modes including sausage modes, kink modes, fluting modes, and torsional (Alfv\'{e}n) modes \citep{roberts2019mhd}.
		 	For sausage modes, the axis of the tube is motionless, while the cross-section area fluctuates with time, resulting in a sausage-like axisymmetric perturbation of the tube boundary. So sausage waves are axisymmetric waves, belonging to magnetoacoustic waves. 
		 	According to the relationship between phase speed $ c_\mathrm{p} $ and sound speed $ c_\mathrm{s} $ in the tube, 
		 	we can divide the sausage \textbf{wave} into \textbf{a} fast mode and slow mode.
		 	The former has a faster phase speed than \textbf{the} sound speed, while the latter has a slower phase speed.
		 	Assumed here and hereafter is that the characteristic speeds inside and outside the tube satisfy the ordering for “photospheric conditions” in the sense of \textbf{Figure} 3 of \citeauthor{roberts1983} (\citeyear{roberts1983}). In particular, $\beta$ inside the tube is assumed to be smaller than unity. This assumption on the internal plasma $ \beta $ is generally accepted for photospheric BPs, \textbf{even though} $\beta$ is difficult to \textbf{measure directly} (e.g. \citealp{shelyag2010,cho2019}).

			Based on ideal MHD equations, \textbf{setting the} \textbf{radial-velocity} perturbation $ v_r = A(r)\cos(kz-\omega t) $, we can obtain \citep[without considering azimuthal components of \textbf{the} velocity perturbance and magnetic field]{freij2016}:
			\begin{equation}\label{bz}
				b_z=\frac{B_0}{\omega}\frac{1}{r}\frac{\partial (rA(r))}{\partial r}\sin(kz-\omega t)\,,
			\end{equation}
			\begin{equation}
		 		\label{p1}
		 		p_1=c_\mathrm{s}^2\rho_1=-\frac{\omega\rho_0c_\mathrm{s}^2}{(c_\mathrm{s}^2k^2-\omega^2)}\frac{1}{r}\frac{\partial(rA(r))}{\partial r}\sin(kz-\omega t)\,,
		 	\end{equation}
	 		where $ b_z $ is the longitudinal magnetic-field perturbation, 
	 		$B_0$ and $\rho_0$ are background magnetic field and density,
	 		$p_1$ is the pressure perturbation, 
	 		and $\rho_1$ is the density perturbation.
		 	
		 	Comparing Equations \ref{bz} and \ref{p1}, we can \textbf{state} that if $c_\mathrm{s}^2k^2-\omega^2>0$ (i.e. the phase speed $c_\mathrm{p}=\omega/k$ is smaller than the sound speed $c_\mathrm{s}$, corresponding to slow sausage mode),
		 	then $b_z$ and $\rho_1$ (or $p_1$) will be 180$^\circ$ out of phase (i.e. in anti-phase).
		 	On the contrary, if $c_\mathrm{s}^2k^2-\omega^2<0$ (i.e. the phase speed is larger than the sound speed, corresponding to \textbf{the} fast sausage mode),
		 	then the perturbations will be in phase.
		 	
		 	Since the magnetic flux in a flux tube is constant, we have $B_0S_0=(B_0+b_{z})(S_0-S_1)$. 
		 	Here, $ S_0 $ and $ S_1 $ represent the unperturbed and perturbed cross-section areas, respectively.
		 	Ignoring $b_{z}S_1$, we can obtain $B_0S_1=-b_{z}S_0$, or 
		 	\begin{equation}
		 		\frac{S_1}{S_0}=-\frac{b_{z}}{B_0}\,.
		 	\end{equation}
		 	So the area perturbation $S_1$ and the longitudinal field perturbation $b_{z}$ are in anti-phase, and we can \textbf{find} the change of \textbf{the} magnetic field from the change of the area.
		 	This was also used \textbf{by} \citeauthor{grant2015} (\citeyear{grant2015}).
		 	
		 	As a result, for \textbf{the} slow sausage mode, fluctuations of area and density are in phase; whereas for \textbf{the} fast sausage mode, the fluctuations are in anti-phase.
		 	These simple phase relations have also been comfirmed by other theoretical investigations \citep{moreels2013cross,moreels2013phase}. 
		 	
		 	Next we will discuss the impact of the opacity effect and associate density with the continuum intensity, 
		 	so that we can obtain the phase relation between fluctuations of intensity and area for fast and slow sausage modes.

		 	\subsection{Phase Relation between Fluctuations of Intensity and Area}
		 	\label{phase}
		 	The TiO data used in our study could be regarded as continuum images (see, e.g. \citealp{leblanc}), so the intensity \textbf{that} we observed
		 	could be written as 
		 	\begin{equation}
		 		I=\int_0^\tau \frac{\sigma T(\tau)^4}{\pi}\mathrm{e}^{-\tau}\mathrm{d}\tau\,,
		 	\end{equation}
		 	where $\sigma$ is the Stefan--Boltzmann constant, and $T(\tau)$ is the local temperature at optical depth of $\tau$. 
		 	Thus the observed intensity fluctuations come primarily from two sources: the change of temperature and the change of optical depth. 
		 	The latter depends on density and temperature \textbf{along} the \textbf{line-of-sight}, 
		 	which is known as the opacity effect \citep{fujimura2009}.
		 	
\begin{figure}
		 		\centerline{\includegraphics[width=\textwidth]{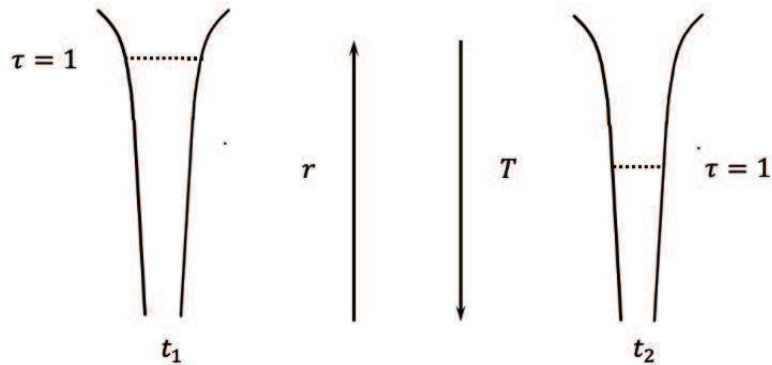}}
		 		\caption{The change of intensity and area caused 
		 			by the opacity effect in a flux tube. 
		 			The dashed lines represent the $\tau=1$ surface, corresponding to 
		 			the BP \textbf{that} we see in the TiO image. From time $t_1$ to $t_2$, 
		 			the surface moves to a lower height. 
		 			The two arrows in the middle indicate that the height increases 
		 			from the bottom up and temperature increases from the top down.}     
		 		\label{sketch1}
		 	\end{figure}

		 	If the oscillations of intensity and area in BPs are caused by the 
		 	opacity effect, then the oscillations
		 	will be in anti-phase. 
		 	In Figure \ref{sketch1}, a BP is shown as a laterally expanding flux tube, and the layer of 
		 	$\tau=1$ is marked with the dashed line. Suppose that from time $t_1$ to $t_2$, 
		 	the $\tau=1$ layer moves to a lower height, resulting in a 
		 	reduced area of the BP. In the deeper layer with a higher 
		 	temperature, the intensity would increase, which means that the 
		 	intensity and area should be in anti-phase. Clearly it is 
		 	contrary to our observation.
		 	
	 	\begin{figure}
	 		\centerline{\includegraphics[width=\textwidth]{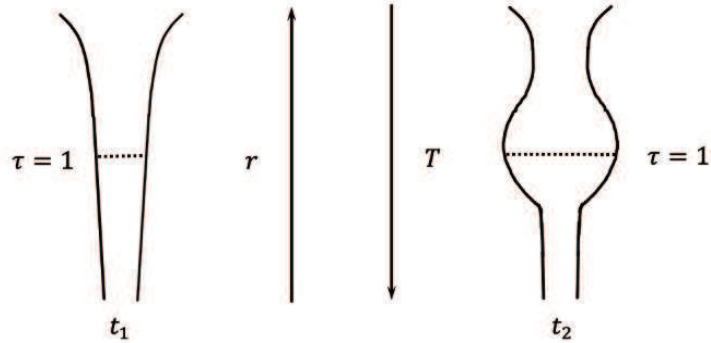}}
	 		\caption{\textbf{The change of intensity and area caused 
	 			by the sausage waves. The $\tau=1$ surface remains the same height from time $t_1$ to $t_2$.  The two arrows in the middle indicate that the height increases 
	 			from the bottom up and temperature increases from the top down.}}     
	 		\label{sketch2}
	 	\end{figure}
		 	
		 	Without considering the opacity effect, we can assume that the \textbf{surfaces that }
		 	we see are at the same height, as shown in Figure \ref{sketch2}. 
		 	Another assumption \textbf{that} we made is that the BP intensity increases from time $ t_1  $ to $t_2 $, and it is caused by the increase of temperature.
		 	The higher temperature will lead to a 
		 	decrease in the absorption coefficient $\kappa_\nu$ at a frequency $ \nu $, which is 
		 	proportional to $[1-\exp\left(-\mathrm{h}\nu/\mathrm{k_B}T\right)]$, where $\mathrm{h}$ and $\mathrm{k_B}$ 
		 	are the Planck constant and Boltzmann constant \citep{rutten1995}. 
		 	In order to ensure 
		 	\begin{equation}
		 		\tau=\int_{r_0}^\infty \kappa_\nu \rho(r) \mathrm{d}r=1\,,
		 	\end{equation}
		 	the density should increase.
		 	Here $\rho(r)$ represents the density at height of $ r $, and $ r_0 $ represents height of the surface \textbf{that} we see.
		 	Our observation shows that the fluctuations of intensity and area are in phase. If the area increases at the same time, the density will be in phase with the area, which is a characteristic of \textbf{the} slow sausage wave (see Section \ref{sausage}).
		 	So the in-phase oscillations of intensity and area that we have 
		 	observed could be possible evidence of slow sausage waves in BPs.
		 	
\begin{figure}
		 		\centerline{\includegraphics[width=\textwidth]{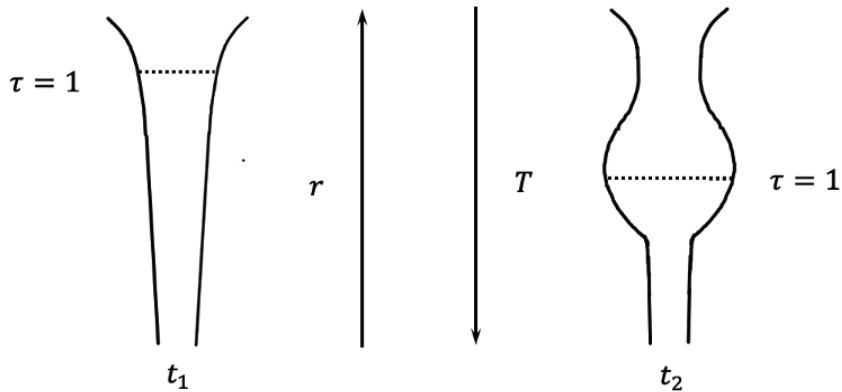}}
		 		\caption{\textbf{The changes of intensity and area caused by both the opacity effect and sausage waves. The $ \tau=1 $ surface moves to a lower height from time $ t_1 $ to $ t_2 $.  The two arrows in the middle indicate that the height increases from the bottom up and temperature increases from the top down.}}     
		 		\label{sketch3}
		 	\end{figure}

		 	However, if we consider both the opacity effect and the presence of sausage modes, 
		 	the result will be different.
		 	In Figure \ref{sketch3}, from $t_1$ to $t_2$, the sausage waves enlarge the cross-section area of the flux tube, and the magnetic field will 
		 	decrease. 
		 	If the opacity effect moves the $\tau=1$ layer to a 
		 	lower height at the same time, we also expect in-phase oscillations 
		 	of intensity and area, but now we cannot distinguish which mode the 
		 	sausage waves are.
		 	The total pressure equilibrium at the same height is given by
		 	\begin{equation}
		 	\label{equili}
		 	p_\mathrm{i}+\frac{B_\mathrm{i}^2}{2\mu_0}=p_\mathrm{e}+\frac{B_\mathrm{e}^2}{2\mu_0}\,,
		 	\end{equation}
		 	where $p_\mathrm{i}$ and $p_\mathrm{e}$ are the internal and external thermal pressure, $B_\mathrm{i}$ and $ B_\mathrm{e} $ are the internal and external magnetic field, and $\mu_0$ is the permeability.
		 	Equation \ref{equili} indicates that the decreasing magnetic field inside the tube will lead to an increasing thermal pressure, which is proportional to $\rho T$. 
		 	If the density $\rho$ increases, the waves will be 
		 	slow mode; otherwise they will be fast mode. 
		 	For specific determination, it may be necessary to introduce 
		 	some appropriate photospheric atmosphere models (e.g. \citealp{fang2006}),
		 	which will be our next step. 
		 	Nevertheless, no matter which mode the waves actually are,
		 	the results \textbf{that} we observed can be seen as \textbf{evidence} for sausage 
		 	waves in BPs.
		 	
	 	\section{Conclusion and Remarks} 
	 	\label{con} 
	 	In this study, we used the TiO images 
	 	obtained from the GST/BBSO on 6 \textbf{August} 2018 to carry out a 
	 	detailed analysis \textbf{of} several isolated photospheric BPs. 
	 	For the four selected BPs, the changes of their area and average 
	 	intensity with time show obvious oscillations, and an in-phase relation.
	 	
	 	In order to interpret these results, we \textbf{presented} a simple
	 	theoretical discussion, ruling out the opacity effect alone and 
	 	identifying the observed oscillations in BPs as 
	 	sausage modes. 
	 	Assuming \textbf{that} the fluctuations are only caused by sausage waves,
	 	we can identify the oscillations in BPs as slow sausage \textbf{modes},
	 	which have been found in some previous observations of magnetic pores
	 	(e.g. \citealp{grant2015,freij2016}). 
	 	Our discovery provides strong support for the existence of 
	 	sausage waves in the photospheric flux tube. 
	 	Although the presence of sausage waves has also been 
	 	observed in magnetic pores or sunspots, 
	 	BPs are better tracers for the flux tubes, 
	 	and \textbf{they} are more ubiquitous in \textbf{the} photosphere. 
	 	Additionally, BPs are much smaller than pores and sunspots, which makes them more easily buffeted by \textbf{surrounding} granular motions.
	 	Since the dissipation of sausage modes is generally believed 
	 	to have an important contribution to the heating of the chromospheric 
	 	atmosphere \citep{grant2015,yu2017,chen2018,keys2018,gilchrist2021}, 
	 	our findings could be helpful for understanding the
	 	chromosphere heating.
	 	
	 	Our \textbf{future} work will first focus on the mode identification by further addressing the opacity effect.
	 	We plan to introduce an appropriate photospheric atmosphere model to figure it out.
	 	\textbf{Futhermore}, our research is still limited to single-channel observations
	 	of \textbf{the} photosphere. We plan to combine more data in different wavelengths or spectral observations 
	 	to study the propagation of sausage waves in BPs to the upper atmosphere.
	 	Moreover, the \textit{Daniel K. Inouye Solar Telescope} (DKIST: \citealp{rast2021}) can provide observations of the photosphere at an unprecedented high resolution.
	 	By using the oncoming data from DKIST, we may find more evidence of MHD waves in BPs and carry out more detailed study. 
	 	
	 	\begin{acks}
	 		We would like to thank Richard Morton, Mingde Ding, and Tom Van Doorsselaere for helpful comments and suggestions. This work has been supported
	 		by the Strategic Priority Research Program of CAS (grant XDA17040507) and NSFC grants 11825301 and 11790304. B. Li was supported by the NSFC grants 11761141002 and 41974200.
	 		BBSO operation is supported by NJIT and US NSF AGS-1821294 grant. GST operation is partly supported by the Korea Astronomy and Space Science Institute, the Seoul National University, and the Key Laboratory of Solar Activities of Chinese Academy of Sciences (CAS) and the Operation, Maintenance and Upgrading Fund of CAS for Astronomical Telescopes and Facility Instruments.
	 	\end{acks}
 	
 		{\footnotesize\paragraph*{Disclosure of Potential Conflicts of Interest} The authors declare that they have no conflicts of interest.}
 	
 	\bibliographystyle{spr-mp-sola}
 	\bibliography{sausage_bibliography}

		\end{article} 
		
	\end{document}